%% file: acl2023.tex
\pdfoutput=1
\documentclass[11pt]{article}

\usepackage[]{ACL2023}

\usepackage{times}
\usepackage{latexsym}

\usepackage[T1]{fontenc}
\usepackage[utf8]{inputenc}

\usepackage{microtype}

\usepackage{inconsolata}

\usepackage{amsmath}
\DeclareMathOperator*{\argmax}{arg\,max}

\usepackage{bbm}

\usepackage{booktabs}
\usepackage{graphicx}

\title{Safer Conversational AI as a Source of User Delight}

\author{Xiaoding Lu \And Aleksey Korshuk \And Zongyi Liu \And Vineet Mudupalli
\AND Christie-Carol Beauchamp \And Thomas Rialan \And William Beauchamp\\
  \AND Chai Research}

\begin{document}
\maketitle
\begin{abstract}
This work explores the impact of moderation on users' enjoyment of conversational AI systems. While recent advancements in Large Language Models (LLMs) have led to highly capable conversational AIs that are increasingly deployed in real-world settings, there is a growing concern over AI safety and the need to moderate systems to encourage safe language and prevent harm. However, some users argue that current approaches to moderation limit the technology, compromise free expression, and limit the value delivered by the technology. This study takes an unbiased stance and shows that moderation does not \textit{necessarily} detract from user enjoyment. Heavy handed moderation does seem to have a nefarious effect, but models that are moderated to be safer can lead to a better user experience. By deploying various conversational AIs in the Chai platform, the study finds that user retention can increase with a level of moderation and safe system design. These results demonstrate the importance of appropriately defining safety in models in a way that is both responsible and focused on serving users.
\end{abstract}

\section{Introduction}
Recent advancements in Large Language Models (LLMs) has lead to the development of highly capable conversational AIs \cite{thoppilan2022lamda}, such as chatGPT. These systems have been rapidly growing in popularity and are being increasingly deployed in real world settings \cite{kung2023performance, huang2022chatbots}. With the prevalence of AI solutions in the real world as tools or as services, there has been huge attention in AI safety and ensuring that developed systems are moderated to be helpful, wholesome and truthful \cite{Detoxify, si2022so, xu2020recipes, manakul2023selfcheckgpt}. There is, however, as well a community of users that believe that systems are being over-moderated, and that moderation limits their freedom of expression and delight in these systems \cite{llanso2020no, wokechatgptarticle}. For example within the entertainment domain, users may wish the conversational AIs to impersonate famous characters, which moderation may restrict them from doing. Further, individual intentions may not always align with the subjective safety standards of the large research institutions developing this technology. It seems important for businesses, and society at large to be willing to discuss what amount and nature of moderation will best serve the needs of the users of this technology.

This paper attempts to take an unbiased stance, and determine the extent to which moderation of conversational AIs influences users' perceived enjoyment of the conversational AI system. We empirically show that safety does not necessarily come at the expense of user enjoyment; models that are moderated to be safer (to a certain extent), even if they ignore user's requests, can lead to a greater user experience. We deploy five conversational AIs in the Chai platform with varying levels of safety and methods to encourage safe language. By considering the downstream user-retention, we find that user-retention can be higher for the moderated systems than for the unmoderated systems. This underscores the importance of a nuanced, light-touch approach to moderation as a way to better serve users.

% Although this may seem like a very obvious finding, we note that trends on social media and entertainment are not necessarily particularly pure and the population today is actually really degenerate, and so these results were surprising to us. 

\section{Related Work}

In recent years, there has been the emergence of a large number of social conversational AIs for chitchat with human users~\citep{yan2022deep, DBLP:journals/corr/abs-2006-16779, DBLP:journals/corr/abs-2001-09977, irvine2023rewarding, choudhary-kawahara-2022-grounding}.\newline 

\noindent\textbf{Unsafe Content}
With the deployment of these powerful conversational AIs in real-world settings, there has been an increasing concern regarding their safety - it is required that systems do not generate harmful or illegal content. \citet{schmidt-wiegand-2017-survey} discuss the scope of unsafe content, covering various aspects of abusive behaviour: hate speech, profanity, malicious intent, hostility, abusive messages and cyber-bullying. There have been extensive efforts in defining thresholds and categories for unsafe responses~\citep{swamy-etal-2019-studying, waseem-etal-2017-understanding, DBLP:journals/corr/MaguJL17, zampieri-etal-2019-semeval, caselli-etal-2020-feel, van-aken-etal-2018-challenges}. A popular categorization~\citep{DBLP:journals/corr/abs-1801-03604, DBLP:journals/corr/abs-2008-12348} for offensive content is: profane content, sexual content, racial content, other hate speech and violent crime. With greater granularity, \citet{2020arXiv200809706Z} outline a more detailed hierarchical taxonomy including concepts such as privacy invasion, jealousy and other forms of malevolent content.\newline 

\noindent\textbf{Design of Safe conversational AI} \label{sec:design}
In conversational AI-human conversations, unsafe content can arise from either the human or the conversational AI. If a human response is identified as harmful, systems will typically return non-committal safe responses or attempt to change the subjects~\citep{cercas-curry-rieser-2019-crowd, DBLP:journals/corr/abs-2008-12348}. However, malicious users can also design undetectable specific adversarial prompts~\citep{DBLP:journals/corr/abs-2005-13170, HILL2015245, DBLP:journals/corr/abs-2004-13637}, where for example a carefully crafted user input can trigger a model to generate offensive content~\citep{wallace-etal-2019-universal}. Adversarial susceptibility of conversational AIs is an open research area. When considering generated conversational AI responses, various approaches have been proposed in literature to mitigate the risk of unsafe content generation. These approaches can be grouped into three main categories~\citep{xu2021recipes}. The first approach is \textbf{data curation}~\citep{DBLP:journals/corr/abs-2004-13637, rashkin-etal-2019-towards}, where models are trained on \textit{safe} data to reduce the likelihood of generating harmful content.  However, such under exposed models struggle to respond to offensive user content and are also susceptible to adversarial prompts~\citep{DBLP:journals/corr/abs-2009-11462}. Alternatively or as an addition to data curation, a safety layer can be introduced through the use of a \textbf{detection system}~\citep{zampieri-etal-2019-semeval, DBLP:journals/corr/abs-1802-00385, dinan-etal-2020-multi, DBLP:journals/corr/abs-1811-12900} that can identify system generated unsafe responses and reject them. Finally, a suite of methods explore \textbf{controlled generation}, where a model is trained to conditionally generate outputs~\citep{DBLP:journals/corr/abs-2009-10855, krause-etal-2021-gedi-generative}, offering the user/designer a level of control on the \textit{offensiveness} of the conversational AI responses. For example \citet{niu-bansal-2018-polite} give a control on model politeness and \citet{nogueira-dos-santos-etal-2018-fighting} show how offensive sentences can be transformed to benign ones.  \newline 

% \noindent\textbf{Safety Evaluation}
% Having designed a process to encourage \textit{safer} chatAIs, it is also necessary to systematically assess how \textit{safe} a designed system is. Although human safety evaluation is popular (\textbf{cite}), the resource costs have led to an increasing use of automated safety evaluation metrics: \textbf{cite}.\newline

\noindent\textbf{Safety and User Engagement}
A typical business is concerned with maximising the \textit{engagingess} of their conversational AI \cite{irvine2023rewarding}, as this translates to a higher level of user retention. There has however been limited research investigating the impact of designing safer models on the level of user engagement. Nevertheless, ~\citet{xu2021recipes} find that there is little correlation between safety and conversational AI engagingness, where human evaluation is used to measure both model safety and engagingness. This work explores the interaction between conversational AI safety and engagingness in greater depth.

\input{method}

\section{Experiments}
\subsection{Set-Up}
% \noindent\textbf{Chatbot Model} In this paper we consider 3 different chatbot systems: \textbf{chatbot-baseline} is a GPT-J 6B \citep{gpt-j} fine-tuned on novels and literature\footnote{\url{https://huggingface.co/hakurei/lit-6B}}. \textbf{chatbot-unconst} is chatbot-baseline further finetuned on real user suggested responses, data which we found contained instances of offensive and hateful responses. While \textbf{chatbot-safe} is the unconstrained model fine-tuned on large quantities of heavily moderated safe data as described in section X. \\

\noindent\textbf{conversational AI Model} In this paper we consider 5 main conversational AI systems: \textbf{chat-ai-baseline-u} is a GPT-J 6B \citep{gpt-j} fine-tuned on novels and literature\footnote{\url{https://huggingface.co/hakurei/lit-6B}}, which was found to contain some unsafe language. \textbf{chat-ai-base-s} is another baseline, where chat-ai-base-u is further fine-tuned on safer literature. Next, we introduce \textbf{chat-ai-safe}, which is the chat-ai-base-s model trained on real user suggested edits on the Chai platform~\footnote{The Chai platform allows for users to edit the response given by the system}, filtered to retain only safe suggestions, as described in Section \ref{sec:design}. For further analysis, we consider \textbf{chat-ai-unsafe}, which again takes the chat-ai-base-s model, but now trains it on all real user suggested edits, which have been found to contain some unsafe language. As a final comparison, we consider a model \textbf{chat-ai-unsafe-mod}, where the chat-ai-unsafe model is deployed with a safety moderation layer to reject user responses detected as unsafe.\\

\noindent\textbf{User Retention} To evaluate the conversational AIs, we deploy them to the Chai Research platform and look at the user retention after 30 days. This is done by creating mutually exclusive cohorts of new users and assigning them to one of the five systems. The 30 day user retention is the number of users that return to the app after 30 days of signing up, and gives a good indication of the user's experience with the conversational AI.\\

\noindent\textbf{Safety Evaluation Dataset} We evaluate the safety of each model as detailed in Section \ref{sec:safe-eval}. The safety score is calculated over a dataset of user-conversational AI conversations, returning the probability of system-generated responses being safe. The safety score (probability) is given by Equation \ref{eqn:safety}. Each model's dataset of system responses consists of 10,000 responses across user conversations, with an average length of \textbf{X} tokens per system response.

\subsection{Results}
Table \ref{tab:comp1} considers the impact of increasing levels of safety on user retention. It is first verified that by further finetuning chat-ai-base-u on safe literature data to produce model chat-ai-base-s does give a safer system, as there is an increase of almost 10\% in average conversational AI responses' safety when deployed on the Chai platform. More encouragingly, there is also observed an increase of 13.8\% in the 30-day user retention on the platform with this safer model. Now, further finetuning the chat-ai-base-s model on proposed user edits, filtered by a safety moderation system, to give the chat-ai-safe model, does indeed encourage the system responses to be significantly safer, as seen by an increase from 60.0\% to 70.2\% in average response safety. Moreover, there is a further 6.9\% increase in the 30-day user retention, suggesting that a safer model can increase user retention.

\begin{table}[htb!]
    \centering
    \begin{tabular}{lcc}
    \toprule
       Model  & Safety (\%) &  Retention (\%$\uparrow$)\\ \midrule
        chat-ai-base-u & 50.5 & - \\
        chat-ai-base-s & 60.0 & +13.7 \\
        chat-ai-safe & 70.2 & +20.6\\
        \bottomrule
    \end{tabular}
    \caption{Model safe-score probability and 30-day user retention relative (percentage increase) to the chat-ai-base-u model.}
    \label{tab:comp1}
\end{table}

Despite the clear trend in Table \ref{tab:comp1}, it can be argued that increased user retention cannot be attributed solely to higher model safety, but also the fact that the safer models have been trained on more data. This proposition is analyzed in further detail in Table \ref{tab:comp2}, where results are presented for the chat-ai-unsafe model, which has also been finetuned on further data. Instead of finetuning chat-ai-base-s on only moderated user suggested edits, all suggested edits are included. A decrease of almost 5\% in the average safety demonstrates that user suggestions often contain threatening language and this encourages the model as a result to generate less safe language. With this decrease in safety there is also a discouraging increase in user retention by 27.6\%, which is greater than the 20.6\% for the chat-ai-safe model. This perhaps shows that it is the finetuning to user suggestions that has more of an impact on user retention than safe model design. 

\begin{table}[htb!]
    \centering
    \begin{tabular}{lcc}
    \toprule
       Model  & Safety (\%) &  Retention (\%$\uparrow$)\\ \midrule
        chat-ai-base-u & 50.5 & - \\
        chat-ai-unsafe & 45.7 & +27.6 \\
        chat-ai-unsafe-mod & 62.2 & +34.5\\
        \bottomrule
    \end{tabular}
    \caption{Model safe-score probability and 30-day user retention relative (percentage increase) to the chat-ai-base-u model.}
    \label{tab:comp2}
\end{table}

\begin{figure}[htb!]
    \centering
    \includegraphics[width=0.4\textwidth]{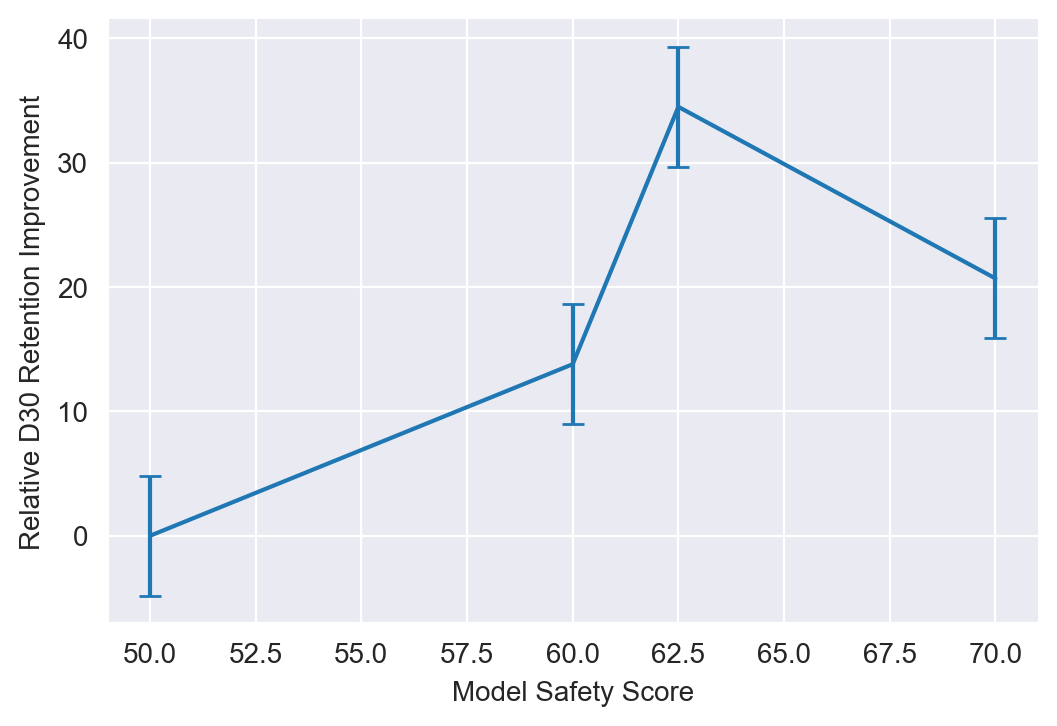}
    \caption{Model Safety probability (\%) against 30-day user retention relative (percentage increase) to the chat-ai-base-u model.}
    \label{fig:plot}
\end{figure}

However, Table \ref{tab:comp2} also shows that model chat-ai-unsafe-mod, a significantly safer version of chat-ai-unsafe (user responses are rejected at deployment time if deemed unsafe), has the greatest user retention, boosting the user retention from +27.6\% to +34.5\%. Therefore, it can be argued that encouraging the model to be safe to a certain extent can increase user retention. Specifically, a safety threshold between 70.2\% and 62.2\% is optimal for maximal user retention, as demonstrated by Figure \ref{fig:plot} summarising the relationship between model safety and user retention for the safer models.

\section{Conclusions}

This work provides empirical evidence that moderation of conversational AI systems can lead to a better user experience without compromising on AI safety. By comparing moderated and unmoderated systems deployed on the Chai chaiAI platform, we find that users often prefer (to an extent) a system that is explicitly trained to encourage safe content. These results have important implications for AI developers and businesses, as they demonstrate that prioritizing safety can not only align with moral duties but also benefit business goals by delivering better products that users prefer. Moving forward, it will be crucial to strike a balance between user entertainment and AI safety, and to continue exploring the impact of moderation on user experience in different settings and domains.

\bibliography{anthology,custom}
\bibliographystyle{acl_natbib}

% \appendix

% \section{Example Appendix}
% \label{sec:appendix}

% This is a section in the appendix.

\end{document}

%% file: method.tex
\section{Method: Design of a safe chat AI}

Modern deep learning-based sequence-to-sequence chat AI systems are typically designed to take as an input a specific user prompt sequence, $\mathbf x$ and a chat history (previous user and system messages), $\mathbf{h}$, to output the response, $\hat{\mathbf y}$ with the highest likelihood, as per the chat AI model, $\mathcal M$,
\begin{equation}
    \hat{\mathbf y} = \argmax_{\mathbf y} \left\{p_{\mathcal M}(\mathbf y| \mathbf x, \mathbf h;\hat\theta) \right\},
\end{equation}
where $\hat\theta$ are the model's trained parameters. These parameters are learnt during training in a standard unsupervised manner, using masked language modeling, where the model predicts the masked tokens. 

\subsection{Design}

The simplest approach to ensure a chat AI system, $\mathcal M$ only generates \textit{safe} sequences is to perform masked language modeling training on only \textit{safe} data, i.e. such that the model is not exposed to unsafe content during training. From a general dataset, $\mathcal D=\{\mathbf y_i\}$, we can extract a safe dataset, $\mathcal D^{(s)}$,
\begin{equation}\label{eqn:safe}
    \mathcal G(\mathbf y) < \epsilon \implies \mathbf y \in\mathcal D^{(s)}, \hspace{2em} \mathbf{y}\in\mathcal D,
\end{equation}
where $\mathcal G$ is a measure of the \textit{threat} of a particular sequence (i.e. a measure of how unsafe the model is), $\mathbf y$ and $\epsilon$ is an agreed threshold of safety. However it is argued that a model under-exposed to unsafe content fails when exposed to harmful user content and may for example just quote the harmful content back. To mitigate the risk of such undesirable behaviour, this work assumes the standard pipeline of pre-training the language model on general data, $\mathcal D$ and then finetuning it on a curated, safe dataset, $\mathcal{D}^{(s)}$. This exposes the model to unsafe content during pre-training but encourages it to only generate safe content after finetuning, i.e. $\mathcal G(\mathbf y) <\epsilon$.

It is expensive and inefficient to use human annotation to define the threat, $\mathcal G$, of any sequence $\mathbf y$ and thus in this work we explore an automated measure to act as a proxy for human perception of model threat. Specifically, a popular and well established moderation tool, OpenAI's moderation endpoint~\footnote{OpenAI moderation tool: \url{https://platform.openai.com/docs/guides/moderation/overview}} is used as a binary threat classifier as a proxy for, $\mathcal G$. The tool also identifies the following groupings: hate, hate/threatening, sexual, sexual/minors, violence, violence/graphic. Any sample in the training set $\mathcal D$ flagged as risk as per these categories, is excluded from the safe dataset, $\mathcal{D}^{(s)}$, as specified in Equation \ref{eqn:safe}.

\subsection{Safety Evaluation} \label{sec:safe-eval}
Having trained a chat AI to be safe, it is necessary to have an independent and systematic measure of the \textit{safety} of generated responses from the chat AI to verify that the design method does truly create \textit{safer} responses. Hence, for a set of chat AI responses, $\{\hat{\mathbf y}\}_{n=1}^N$, we can calculate a safety score, $\mathcal S$ defined as:
\begin{equation} \label{eqn:safety}
    \mathcal S = 1 - \frac{1}{N}\sum_{n=1}^N\tilde{\mathcal G}(\hat{\mathbf y}),
\end{equation}
where $\tilde{\mathcal G}$ represents the threat score normalized to lie between 0 and 1, to behave as a probability. Once again, it is expensive and inefficient to use human evaluation at scale for the threat measure, $\mathcal G$ and thus this work uses an automated proxy measure of threat at inference time. To reduce bias towards the design method, it is necessary to use a different (from open AI's moderation tool) proxy function for $\mathcal G$. Hence, this work uses a standard deep-learning based model, $\mathcal Q$, trained in a supervised manner on pre-curated public safety dataset (binary classification task to identify unsafe and safe samples)~\footnote{Model used for evaluation of safety: \url{https://huggingface.co/unitary/toxic-bert}}, as a proxy for threat classification at inference time,
\begin{equation}
    \mathcal G(\mathbf y)\approx P_{\mathcal Q}(\mathbf y).
\end{equation}